\documentclass[12pt]{article}
\usepackage{amsfonts}
\usepackage{pstricks}
\usepackage{pst-node}

 \usepackage{fancyhdr}
 \usepackage{graphicx}
 \usepackage{fancybox}
 \usepackage{fancyvrb}
\begin{document}

\def\ba{\begin{eqnarray}}
\def\ea{\end{eqnarray}}

\begin{titlepage}
\title{{\bf Nonmetricity and torsion induced by dilaton gravity in two dimension}}
\author{ M. Adak  \\
 {\small Department of Physics, Faculty of Arts and Sciences, Pamukkale University,}\\
{\small 20100 Denizli, Turkey} \\ {\small madak@pamukkale.edu.tr}}
\date{\today }
\maketitle

  \thispagestyle{empty}

\begin{abstract}
We develop a theory in which there are couplings amongst Dirac
spinor, dilaton and non-Riemannian gravity and explore the nature
of connection-induced dilaton couplings to gravity and Dirac
spinor when the theory is reformulated in terms of the Levi-Civita
connection. After presenting some exact solutions without spinors,
we investigate the minimal spinor couplings to the model and in
conclusion we can not find any nontrivial dilaton couplings to
spinor.
\end{abstract}
\end{titlepage}

\section{\large Introduction}

It is worthwhile to study dilaton gravity theories because they
are connected with black holes in effective string models and may
be related to higher dimensional gravity theories for some special
choices of the dilaton and matter couplings. Two ways can be
followed in order to write nontrivial dilaton gravity models in
Riemannian spacetimes. In the first method, a simple Lagrangian
$D$-form, $ \mathbb{L}$, with Levi-Civita connection in $D>2$
dimension is guessed and the new Lagrangian $(D-1)$-form, $L$, is
obtained via Kaluza-Klein dimensional reduction procedure in which
the Levi-Civita connection is calculated uniquely from a
particular $D$-dimensional metric, $\mathbb{G}$, including
$(D-1)$-dimensional metric, $g$, gauge potentials, $A$,
(components of vector fields) and dilatons, $\phi$, (scalar
fields) \cite{tdereli82}, \cite{madak04}:
 \ba
  \stackrel{(D)}{\mathbb{G}}= \stackrel{(D-1)}{g}
  + f^2(\phi) A\otimes A + f^2(\phi) (dy \otimes A +A \otimes dy )
   + f^2(\phi) dy \otimes dy \quad \rightarrow \quad
  \stackrel{(D)}{ \mathbb{L}} = \stackrel{(D-1)}{ L} \wedge dy \nonumber
 \ea
where $y$ is a coordinate function living only in $D$-dimensional
manifold. This procedure can be applied successively several times
for lower dimensions.

In the second approach, non-Riemannian geometry in which the full
connection, ${\Lambda^a}_b$, contains Levi-Civita, ${\omega^a}_b$,
torsion, $T^a$, and nonmetricity, ${Q^a}_b$, contributions is
used. Here variational calculation with constraints on
nonmetricity and torsion is needed. After writing a non-Riemannian
Lagrangian $D$-form, firstly the full connection is calculated via
the constraint equations as Levi-Civita plus dilaton terms,
$\Lambda \approx \omega + \phi $, and then by inserting the solved
connection to the other variational equations, the equations with
the standard Levi-Civita connection are obtained. Now one repeats
the operations in reverse order, that is, firstly non-Riemannian
Lagrangian is decomposed by inserting the solved connection, and
then the field equations are derived from the new Lagrangian
\cite{tdereli94}. In this approach, it is seen that the theory can
be rewritten in terms of Levi-Civita and thus torsion and
nonmetricity tensors are interpreted as matter induced couplings
for Riemannian gravity. This scheme may be applied to supergravity
in which models generally includes complicated matter couplings
with respect to Levi-Civita connection and these may have a much
tidier form when they are reformulated in terms of the full
connection with torsion and nonmetricity.

In this paper, by adding nonzero torsion we generalize the paper
\cite{tdereli94}, in which authors showed how theories of dilaton
gravity can be constructed in terms of a torsion-free nonmetric
connection and found the corresponding renormalization of the
connection induced couplings when the theory is reformulated in
terms of the Levi-Civita connection. Then we discuss some
solutions and finally investigate whether there are nontrivial
dilaton couplings to spinor by treating Dirac Lagrangian.

We make use of the following conventions and notations. The
signature of the 2-dimensional metric is assumed to be $(-,+)$.
While Latin indices, $a,b, \cdots = 0,1$, label the orthonormal
frame components, Greek indices, $\alpha , \beta , \cdots =
\hat{0}, \hat{1}$, denote the coordinate frame components. Along
with the orthonormal co-frame 1-forms $e^a$ and the exterior
product $\wedge$, we will use the short hand notation $e^a \wedge
e^b = e^{ab}$. Furthermore, we define the spacetime orientation in
terms of the Hodge dual such that $* 1= e^{01}$ is the volume
$2$-form.

\section{Dilaton gravity in non-Riemannian geometry}

A two dimensional spacetime consists of a differentiable manifold
$M$ equipped with a Lorentzian metric $ g $  and a linear
connection $\nabla$ that defines parallel transport of vectors (or
tensors) and more generally spinors. Given  an orthonormal basis
$\{ X_a \}$, the metric reads
 \ba
      g = \eta_{ab}e^a \otimes e^b = -e^0 \otimes e^0 + e^1 \otimes e^1 \label{metric}
 \ea
where $ e^a $ is the orthonormal co-frame such that
 \ba
     \imath_a e^b = e^b(X_a)=\delta^b_a \; .
 \ea
Here $\imath_{X_a} \equiv \imath_a$ denotes interior product. The
nonmetricity 1-forms, torsion 2-forms and curvature 2-forms are
defined by the Cartan structure equations
 \ba
   2Q_{ab} &:=& -D\eta_{ab} = \Lambda_{ab} +\Lambda_{ba} \; \; , \label{nonmet}\\
   T^a &:=& De^a = de^a + {\Lambda^a}_b \wedge e^b \; \; , \label{torsion}\\
  {R^a}_b &:=& D{\Lambda^a}_b := d{\Lambda^a}_b +{\Lambda^a}_c \wedge {\Lambda^c}_b \label{curva}
 \ea
where $ d \; ,\; D$ denote the exterior derivative and the
covariant exterior derivative, respectively. The linear connection
$\nabla$ is determined by the connection 1-forms ${\Lambda^a}_b$
which can be decomposed in a unique way according
to~\cite{tdereli96}:
 \ba
  {\Lambda^a}_b = {\omega^a}_b + {K^a}_b + {q^a}_b + {Q^a}_b \;  \label{connec}
 \ea
where $ {\omega^a}_b $ are the Levi-Civita connection 1-forms:
 \ba
     {\omega^a}_b \wedge e^b =-de^a \quad \mbox{or} \quad
     2 \omega_{ab} = - \imath_a (de_b) + \imath_b (de_a) + \imath_a \imath_b (de_c) e^c
 \ea
$ {K^a}_b $ are the contortion 1-forms:
 \ba
     {K^a}_b \wedge e^b =T^a \quad \mbox{or} \quad
     2 K_{ab} =  \imath_a (T_b) - \imath_b (T_a) - \imath_a \imath_b (T_c) e^c
 \ea
and $ {q^a}_b $ are the anti-symmetric tensor 1-forms:
 \ba
     q_{ab} = -(\imath_a Q_{bc}) e^c
        + (\imath_b Q_{ac}) e^c \; .\label{antisy}
 \ea

The literature on the most general non-Riemannian gravity in
two-dimension may be found in \cite{yobukhov04} in which in spite
of the inclusion of nonmetricity and torsion together, the author
has not discussed the dilaton couplings to his model. Furthermore,
while the simplest model of dilaton gravity can be written as the
coupling of a dilaton scalar $\phi$ to the curvature scalar in
two-dimension, we will develop a theory with a connection
including nonmetricity determined by (same as \cite{tdereli94})
 \ba
{Q^b}_a =\delta^b_a ( k d\!\phi + l  \, *\!d\!\phi ) \;.
\label{non-metricity}
 \ea
and torsion determined by
 \ba
     T^a = e^a  \wedge ( p d\phi + q *d\phi ) \label{Tconst}
 \ea
where $k,l,p$ and $q$ are fundamental coupling constants. In
section $6$ of the reference \cite{mokatanaev02}, the author
proved that a general two-dimensional dilaton gravity is
equivalent to two-dimensional gravity with torsion in terms of the
first order Hamiltonian formulation. We notice that our {\it ad
hoc} anzatz for torsion is motivated by the equations of motion of
that analysis. Our theory is based on the Lagrangian $2$-form
 \ba
   L &=& \frac{1}{2} \phi^2 {R^a}_b \wedge *{e_a}^b
 + \frac{\alpha}{2}d\phi \wedge * d\phi + \frac{\beta}{2} \phi^2 *1
 + \frac{\mu}{4}{Q^a}_b \wedge * {Q^b}_a + \frac{\nu}{2} T^a \wedge *T_a \nonumber \\
 & & + {\rho^a}_b \wedge ({Q^b}_a - k\delta^b_a d\phi -l \delta^b_a *d\phi)
     + \lambda_a (T^a-pe^a \wedge d\phi -qe^a \wedge *d\phi) \label{lagrange}
 \ea
where $\alpha, \beta, \mu, \nu$ are coupling constants,
${\rho^a}_b$ symmetric Lagrange multiplier $1$-forms constraining
the nonmetricity to (\ref{non-metricity}) and $\lambda_a$ Lagrange
multiplier $0$-forms ensuring the torsion constraint
(\ref{Tconst}). In the following $\rho = {\rho^a}_a$ and $\lambda
= \lambda_a e^a$. Variations with respect to ${\Lambda^a}_b$,
$\phi$ and $e^a$ give the field equations:
 \ba
 \frac{1}{2} d\phi^2 \wedge *{e_a}^b +\frac{\phi^2}{2} (2Q^{bc} \wedge *e_{ac} -Q \wedge *{e_a}^b)
  + \frac{\mu}{2}*{Q^b}_a - {\rho^b}_a + \lambda_a e^b +\nu e^b *T_a &=&0 \label{multipliers} \\
 \phi {R^a}_b \wedge *{e_a}^b - \alpha d*d\phi + \beta \phi *1
  - k d\rho + l d*\rho -p d\lambda + q d*\lambda &=& 0 \label{phi}\\
  -\frac{\alpha}{2} \tau_a{[\phi]} + \frac{\beta}{2} \phi^2*e_a - \frac{\mu}{4} \tau_a{[Q]}
   + l [ (\imath_a d\phi) *\rho + (\imath_a *d\phi) \rho ] +\nu D*T_a \quad \quad \quad \quad { } & &\nonumber \\
  - \frac{\nu}{2} (\imath_a T^b) *T_b  + D\lambda_a -p \lambda_a d\phi -q \lambda_a *d\phi
+q[(\imath_a d\phi) *\lambda + (\imath_a *d\phi) \lambda] &=&0
\label{e^a}
 \ea
where the stress forms are
 \ba
     \tau_a[\phi] = (\imath_a d\phi ) *d\phi + (\imath_a *d\phi)d\phi
 \ea
while
 \ba
    \tau_a[Q] = ( \imath_a {Q^b}_c ) *{Q^c}_b + ( \imath_a *{Q^c}_b ) {Q^b}_c \; .
 \ea

We can solve the multipliers by first inserting
(\ref{non-metricity}) and (\ref{Tconst}) into (\ref{multipliers})
 \ba
     \frac{1}{2} d\phi^2 \wedge *{e_a}^b + \frac{\mu}{2} \eta_{ab}(k*d\phi +l d\phi)
     -\nu e_b \wedge [p (\imath_a *d\phi +q (\imath_a d\phi)] + \lambda_a e_b  -
     \rho_{ab}=0 \label{Mba}\; .
 \ea
Now after contracting (\ref{Mba}) with $\imath^a$ and with
$\imath^b$ we subtract these equations side by side after
relabeling the indices in one of them
 \footnote{Last two terms of equation (44) in Ref.\cite{tdereli94} were mistyped.
Fortunately, this mistyping does not change the results.},
 \ba
     \lambda_a = \imath_a (2\phi *d\phi +\nu p *d\phi +\nu q d\phi) \label{lambdaa}
 \ea
and
 \ba
    \lambda = 2\phi *d\phi +\nu p *d\phi +\nu q d\phi \; .
 \ea
Inserting (\ref{lambdaa}) in (\ref{Mba}) gives
  \ba
     \rho = \mu l d\phi +\mu k *d\phi +2\phi *d\phi \; .
  \ea
Putting these results for the multipliers into (\ref{phi}) and
(\ref{e^a}) yields the field equations for the dilaton and metric
from the Lagrangian (\ref{lagrange}).

\section{Reduction to a theory with the Levi-Civita connection}

By inserting (\ref{non-metricity}) and (\ref{Tconst}) in
(\ref{connec}) the full connection 1-forms can be written in terms
of the Levi-Civita 1-forms and dilaton contributions
 \ba
    {\Lambda^a}_b = {\omega^a}_b + {\epsilon^a}_b [ (k+p) *d\phi
      +(l+q) d\phi] + \delta^a_b (k d\phi + l *d\phi )
      \label{Lambdaab}
 \ea
In the same manner the curvature 2-forms can be decomposed as
follows
 \ba
     {R^a}_b (\Lambda)={R^a}_b(\omega ) + [ (k+p) {\epsilon^a}_b + l \delta^a_b] d*d\phi \label{Rlevi}\; .
 \ea
Thus Einstein-Hilbert term takes the form
 \ba
     {R^a}_b \wedge * {e_a}^b = {R^a}_b(\omega ) \wedge * {e_a}^b - 2(k+p) d*d\phi
 \ea
and similarly
 \ba
    \tau_a[Q] &=& 2(k^2+l^2)\tau_a[\phi] +4kl *\tau_a[\phi] \\
 (\imath_a T^b) *T_b &=& (p^2-q^2) (\imath_a d\phi)*d\phi -(p^2-q^2) (\imath_a *d\phi)d\phi \\
 D*T_a &=& D(\omega)*T_a +(pk+ql) \tau_a[\phi] + (kq+pq+pl)*\tau_a[\phi]\nonumber  \\
              & & \quad \quad \quad \quad  \quad \quad \quad \quad
                     \quad \quad \quad \quad
                     + p^2 (\imath_a d\phi)*d \phi + q^2 (\imath_a *d\phi)d \phi \\
 D\lambda_a &=& D(\omega )\lambda_a - (2k\phi +\nu pk +\nu ql) \tau_a[\phi]
             - (2l \phi +\nu pq +\nu pl +\nu qk) *\tau_a[\phi] \nonumber \\
        & & \quad \quad \quad \quad \quad \quad -(2p\phi +\nu p^2) (\imath_a d\phi)*d \phi
        - \nu q^2 (\imath_a *d\phi)d \phi -2q\phi (\imath_a d\phi)d \phi
 \ea
where $D(\omega)$ denotes the covariant exterior derivative with
respect to Levi-Civita. By using these expressions in (\ref{phi})
and (\ref{e^a}) we obtain that the field equations for the dilaton
and metric are rewritten just in terms of the Levi-Civita
connection.
 \ba
    \phi {R^a}_b(\omega ) *{e_a}^b - 4(k+p)\phi d*d\phi
     - 2(k+p) d\phi \wedge *d\phi \quad \quad \quad \quad \quad \quad \quad \quad  \quad & &\nonumber \\
     -[\alpha + \mu (k^2-l^2) +\nu (p^2-q^2)] d*d\phi + \beta \phi *1 &=&0 \label{phi2}\\
    D(\omega )(\imath_a *d\phi^2) + \frac{\beta }{2}\phi^2 *e_a -2(k+p)\phi\tau_a[\phi ]
     - \frac{1}{2} [ \alpha +\mu (k^2-l^2) +\nu (p^2-q^2)] \tau_a[\phi ] &=& 0 \label{e^a2}
  \ea
Now we decompose the followings by using the results above
 \ba
   \phi^2 {R^a}_b \wedge * {e_a}^b &=& \phi^2{R^a}_b(\omega ) \wedge * {e_a}^b + 4(k+p)\phi d\phi \wedge *d\phi \\
     {Q^a}_b \wedge *{Q^b}_a &=& 2(k^2-l^2)d\phi \wedge *d\phi \\
    T^a \wedge *T_a &=& (p^2-q^2)d\phi \wedge *d\phi \; .
 \ea
Inserting these results in (\ref{lagrange}) yields the new
Lagrangian with Levi-Civita connection.
  \ba
   L&=& \frac{1}{2} \phi^2 {R^a}_b(\omega) \wedge *{e_a}^b
          +\frac{\beta}{2} \phi^2*1 +\lambda_a T^a + {\rho^a}_b \wedge {Q^b}_a \nonumber \\
    & &+ \{ 2(k+p)\phi + \frac{1}{2} [ \alpha +\mu (k^2 - l^2) +\nu (p^2-q^2) ] \} d\phi \wedge
    *d\phi \label{lagrange2}
 \ea
We have also verified that the new Lagrangian gives rise to the
field equations (\ref{phi2}) and (\ref{e^a2}). We point out the
observation of how the nonmetricity and torsion tensors cause to
kinetic terms of the scalar field and then rescale the stress
forms of the scalar field and finally yield an original derivative
scalar interaction for $k+p\neq 0$. Besides, we notice that our
model is a subcase of the one discussed in \cite{dgrumiller02},
which is formulated just in terms of Rimennian geometry, under the
following definitions of the potentials
 \ba
      U(X) &=& -\frac{k+p}{\sqrt{X}} - \frac{\alpha + \mu (k^2-l^2) +\nu
      (p^2-q^2)}{4X} \\
      V(X) &=& \frac{\beta}{2} X
 \ea
where our dilaton is related to theirs via $\phi^2 = X$.
Furthermore, it seems that one can extend our model by allowing
$k,l,p,q,\alpha,\beta,\mu,\nu$ to be not just constants, but
rather arbitrary functions of the dilaton field. Then,
(\ref{lagrange2}) will be essentially equivalent to (1.1) in ref.
\cite{dgrumiller02}. For the further references concerning the
dilaton gravity theories and their applications to black hole
physics and string models one can consult to
ref.\cite{ewitten91}-\cite{vdalfaro}.

 \section{Discussion on solutions}

The model is dependent of the eight real parameters $\{ \alpha ,
\beta , \mu , \nu , k ,l, p, q  \}$ more general than
\cite{tdereli94}. Firstly we investigate the static solutions:
 \ba
     e^0 = F(x) dt \quad , \quad \quad e^1= \frac{dx}{F(x)} \quad , \quad
     \quad \phi = \phi(x)
 \ea
In this case, (\ref{phi2}), zeroth  and first components of
(\ref{e^a2}) read explicitly; respectively
 \ba
    \phi {(F^2)}'' + \{ 2(k+p){(\phi^2)}' + [ \alpha +\mu (k^2-l^2)
      + \nu (p^2-q^2)]\phi' \} {(F^2)}' \quad \quad \quad \quad \quad& &\nonumber \\
 + \{ 2(k+p){(\phi')}^2 + 4(k+p)\phi \phi'' + [ \alpha +\mu (k^2-l^2)
            + \nu (p^2-q^2)]\phi'' \} F^2 &=& \beta \phi \label{phistatic} \\
 \frac{1}{2}{(\phi^2)}' {(F^2)}' +  [ 2{(\phi')}^2 + 2 \phi \phi'' ] F^2
   \quad \quad \quad \quad \quad \quad  \quad \quad \quad \quad \quad  \quad
  \quad \quad & & \nonumber \\
 - \{ 2(k+p) \phi + \frac{1}{2}[\alpha + \mu (k^2-l^2) +\nu (p^2 -q^2)] \}
                    {(\phi)'}^2 F^2 &=& \frac{\beta}{2}\phi^2 \label{0comstatic} \\
  \frac{1}{2}{(\phi^2)}' {(F^2)}' + \{ 2(k+p) \phi + \frac{1}{2}[\alpha + \mu (k^2-l^2) +\nu (p^2 -q^2)]\}
   {(\phi)'}^2  F^2 &=& \frac{\beta}{2}\phi^2  \label{1comstatic}
 \ea
where prime denotes the derivative with respect to $x$. First by
adding (\ref{0comstatic}) and (\ref{1comstatic})
 \ba
     {(\phi')}^2 F^2 = \frac{\beta}{2} \phi^2 - \phi \phi'' F^2 -
     \frac{1}{2}{(\phi^2)}'{(F^2)}'
 \ea
and then by inserting this in (\ref{phistatic}) one obtains
  \ba
    \phi {(F^2)}'' + \{ (k+p){(\phi^2)}' + [ \alpha +\mu (k^2-l^2)
      + \nu (p^2-q^2)]\phi' \} {(F^2)}' \quad \quad \quad \quad \quad& &\nonumber \\
 + \{ 2(k+p) \phi \phi'' + [ \alpha +\mu (k^2-l^2)
            + \nu (p^2-q^2)]\phi'' \} F^2 &=& \beta \phi \; .
 \ea
This is the our most general equation and it seems impossible to
solve generally it. Therefore, we look at some special cases.

\begin{center}
 { \it Special Case:1 } $k+p=0$ and $\alpha +\mu (k^2-l^2) + \nu (k^2-q^2) =4$
\end{center}
Here we first drop all the nonlinear terms in $\phi$ and find the
solution
 \ba
     \phi (x) = e^{c_1x} \quad \quad , \quad \quad F^2(x) = \frac{\beta +4 c_1^2 c_2
     \phi^{-2}(x)}{4c_1^2}
 \ea
where $c_1 , c_2$ are constants. This is known as {\it dilaton
black hole} \cite{ewitten91}. Our constraints on the parameters
comprise the related cases discussed in \cite{tdereli94} and more.

\begin{center}
 { \it Special Case:2 } $k+p=0$ and $F^2=1$
\end{center}
One has to pay attention for interpreting this case. At first
glance it seems that we work in Minkowski spacetime. This is true
if we write the theory in terms of only Levi-Civita without
thinking of nonmetricity and torsion like (\ref{lagrange2}), but
if we take nonmetricity and torsion into account the situation
becomes quite different. In this case, although Riemannian
curvature is zero, the non-Riemannian curvature in general is not
zero because of (\ref{Rlevi}) as long as $l \neq 0$ and thus we
are still in a spacetime with curvature, torsion and nonmetricity.
Technically speaking; while the worldlines of test particles
coincide with the geodesics in the limit of $k=l=p=q=0$ couplings,
autoparallels of the full connection deviate from geodesics in the
cases of nonzero $k,l,p,q$ constants, and then we interpret these
deviations as nonmetricity and torsion. Now we write down the
solution as
 \ba
     F^2 (x)=1 \quad \quad , \quad \quad
     \phi (x) =c_1 e^{\sqrt{-\frac{\beta}{c_3}}} + c_2 e^{-\sqrt{-\frac{\beta}{c_3}}}
 \ea
where $c_1 , c_2$ are arbitrary constants and $c_3 = \alpha +\mu
(k^2-l^2) +\nu (k^2 -q^2)$. Here it is interesting to observe that
while if $\beta$ and $c_3$ have the same sign, the solution is
periodic, it is hyperbolic when they have the opposite signs.

\begin{center}
 { \it Special Case:3 } $\alpha +\mu (k^2-l^2) + \nu (p^2 - q^2)=0$ and $F^2=1$
\end{center}
In this case we obtain the solution
 \ba
    F^2 (x)=1 \quad \quad , \quad \quad
     \phi (x) = - \frac{1}{k+p} + c_1 e^{\sqrt{-\frac{\beta}{2}}} + c_2 e^{-\sqrt{-\frac{\beta}{2}}}
 \ea
for arbitrary constants $c_1, c_2$. We point out that the solution
depends crucially  on $k+p$ and periodicity is dependent of the
sign of $\beta$. Here while we are again in Minkowski apacetime in
the Riemannian sense, we are in a curved spacetime with
nonmetricity and torsion in the non-Riemannian sense.

Finally we examine the cosmological type solutions:
 \ba
     e^0 = dt \quad , \quad \quad e^1 = F(t)dx \quad , \quad
     \quad \phi = \phi(t)
 \ea
This time, (\ref{phi2}), zeroth  and first components of
(\ref{e^a2}) read explicitly;
 \ba
    2 \phi \ddot{F} + \{ 4(k+p) \phi \dot{\phi} + [ \alpha +\mu (k^2-l^2)
      + \nu (p^2-q^2)] \dot{\phi} \} \dot{F} \quad \quad \quad \quad \quad \quad \quad \quad \quad \quad& &\nonumber \\
 + \{  4(k+p)\phi \ddot{\phi} + [ \alpha +\mu (k^2-l^2)
            + \nu (p^2-q^2)] \ddot{\phi} +2(k+p) {(\dot{\phi})}^2  + \beta \phi \} F &=&0 \label{phicosmo} \\
 2\phi \dot{\phi} \dot{F} + \{ 2(k+p) \phi + \frac{1}{2}[\alpha + \mu (k^2-l^2) +\nu (p^2 -q^2)] \}
                    (\dot{\phi})^2 F + \frac{\beta}{2}\phi^2 F &=&0  \label{0comcosmo} \\
  2 \phi \ddot{\phi} + 2 (\dot{\phi})^2 - \{ 2(k+p) \phi
  + \frac{1}{2}[\alpha + \mu (k^2-l^2) +\nu (p^2 -q^2)] \} (\dot{\phi})^2 + \frac{\beta}{2}\phi^2 &=&0  \label{1comcosmo}
 \ea
where dot denotes the derivative with respect to $t$.  After
multiplying (\ref{1comcosmo}) by $F$, addition of the equations
(\ref{0comcosmo}) and (\ref{1comcosmo}) yields
 \ba
     \phi \dot{\phi} \dot{F} = -[\phi \ddot{\phi} + (\dot{\phi})^2 +\frac{\beta}{2}\phi^2]F
 \ea
and then inserting this into (\ref{phicosmo}) gives
 \ba
    & & 2\phi \ddot{F} + [\alpha + \mu (k^2-l^2) +\nu (p^2 -q^2)] \dot{\phi} \dot{F} \nonumber \\
    & & \quad \quad \quad + \{[\alpha + \mu (k^2-l^2) +\nu (p^2 -q^2)]\ddot{\phi}
     -2(k+p) (\dot{\phi})^2 - 2(k+p)\beta \phi^2 +\beta \phi \} F
     =0 \; .
 \ea
If we take $\alpha + \mu (k^2-l^2) +\nu (p^2 -q^2) =6$, then
 \ba
     \phi(t)= \frac{\beta}{4(k+p)}t^2  \quad \quad ,  \quad \quad
     F(t) =c \frac{e^{\frac{1}{4}\beta t^2}}{t^3}
 \ea
for the integration constant $c$. Finally, for $k+p=0$ we have the
inflationary type solution
 \ba
 \phi (t) = e^{-\frac{c_1}{2}t} \quad \quad ,  \quad \quad
 F(t) = e^{c_1 t}
 \ea
where
 \ba
     c_1 = \pm \sqrt{\frac{4\beta}{\alpha + \mu (k^2-l^2) +\nu (k^2
     -q^2)-8}} \; .
 \ea

 \section{Discussion on spinor couplings to the model}

In this section we investigate if there are nontrivial couplings
of spinor to the dilaton field. We use the following Dirac
matrices as the generators of the Clifford algebra
${\mathcal{C}}\ell_{1,1}$
 \ba
   \gamma_0  &=&  \left (
                 \begin{array}{cc}
                                   0   & 1  \\
                                   -1  & 0
                 \end{array}
          \right )             \; , \;\;\;
   \gamma_1  =   \left (
                 \begin{array}{cc}
                                   0    &  1  \\
                                   1 &  0
                 \end{array}
          \right ).  \;  \label{dirmat}
 \ea
which they satisfy the anticommutation relation
 \ba
     \{\gamma_a , \gamma_b \} = 2 \eta_{ab}
 \ea
and the commutation relation
 \ba
     [\gamma_a , \gamma_b] = 4 \sigma_{ab}
 \ea
where $\sigma_{ab}$ are the generators of the Lorentz group. Since
we use the $2 \times 2$ matrix representations of
${\mathcal{C}}\ell_{1,1}$, we represent the Dirac spinor $\psi$ as
a 2-component complex valued column matrix whose covariant
exterior derivative is given explicitly by \cite{madak03}
 \ba
  D\psi = d\psi + \frac{1}{2}\Lambda^{[ab]}\sigma_{ab} \psi
             + \frac{1}{4}Q \psi \label{covderpsi}
 \ea
where the Weyl 1-form $ Q = {Q^a}_a $. Finally, the curvature of
the spinor bundle is given by
 \ba
    D(D\psi ) = \frac{1}{2}R^{[ab]}\sigma_{ab} \, \psi
    -\frac{1}{2}{Q^a}_c \wedge Q^{cb}\sigma_{ab} \, \psi
    + \frac{1}{4}dQ \, \psi \; . \label{spinorcurve}
 \ea
In the Weyl geometry, i.e. $Q_{ab} = \eta_{ab} d \phi $ where
$\phi$ any scalar field, the last two terms vanish.

The Dirac Lagrangian 2-form is written in terms of
${\mathcal{C}}\ell_{1,1}$-valued 1-forms $ \gamma = \gamma^a e_a $
and  the inverse of the Compton wavelength $M=\frac{mc}{\hbar}$ as
follows
 \ba
   L_D = \frac{i}{2}(\overline{\psi} * \gamma \wedge D\psi
 + \overline{D\psi} \wedge * \gamma \psi )+ iM\overline{\psi}\psi *1 \label{diraclag}
 \ea
where the Dirac adjoint of a spinor is defined $\overline{\psi} =
\psi^\dag \gamma_0 $. While variation with respect to
${\Lambda^a}_b$ gives no contribution to (\ref{multipliers}) since
$\gamma_5 \gamma_a +\gamma_a \gamma_5 =0$ with the definition
$\gamma_5 = \gamma_0 \gamma_1$, one obtains the following
contribution to (\ref{e^a}) after $e^a$ variation
 \ba
     \tau_a[\psi] =  \frac{i}{2}[\overline{\psi} (\imath_a * \gamma) \wedge D\psi
 - \overline{D\psi} \wedge (\imath_a * \gamma) \psi ]+ iM\overline{\psi}\psi
 *e_a \; .
 \ea
Finally, $\overline{\psi}$ variation gives the Dirac equation:
 \ba
   * \gamma \wedge D\psi + M\psi *1 - \frac{1}{2} * \gamma \wedge T \psi
   - \frac{3}{4} * \gamma \wedge Q \psi - \frac{1}{2} \gamma_a Q^{ab} \wedge *e_b \psi =0
\label{psibar}
 \ea
where $T=\imath_a T^a$. When we insert (\ref{Lambdaab}) into
(\ref{covderpsi}), we obtain the decomposed covariant derivative
of spinor
 \ba
    D\psi = D(\omega )\psi - \frac{1}{2} [ (k+p) * d\phi +(l+q) d\phi )\gamma_5 \psi
     + \frac{1}{2}(k d\phi + l * d\phi )\psi \; . \label{Dpsi}
 \ea
Substitution of this result into (\ref{diraclag}) yields
  \ba
   L_D = \frac{i}{2}(\overline{\psi} * \gamma \wedge D(\omega)\psi
 + \overline{D(\omega )\psi} \wedge * \gamma \psi )+ iM\overline{\psi}\psi *1
 \ea
which gives rise to the variational field equation
 \ba
  * \gamma \wedge D(\omega )\psi + M\psi *1 =0 \; .
 \ea
We have also verified that this equation is obtained when
(\ref{Dpsi}) is put into (\ref{psibar}). Thus we showed that the
interpretations of matter-dilaton couplings of nonmetricity and
torsion do not produce any nontrivial spinor-dilaton couplings in
this model.

 \section{Conclusions}

We have generalized the theory \cite{tdereli94}, in which the
authors developed a 2-dimensional theory of dilaton gravity in
terms of a zero-torsion and nonmetric-compatible connection
determined by a two parameter nonmetricity tensor, by adding
non-zero torsion determined by a further two parameters. After
solving the Lagrange multipliers algebraically from the field
equations we discussed some exact classical solutions for static
and cosmological types. For more classical solutions of these
kinds of models one can consult \cite{dgrumiller02} in that the
authors use a different gauge from the diagonal gauge. Basically,
this kind of torsion contribution to the model shifts simply the
coupling constants. Besides, we rewrote the theory in terms of the
Levi-Civita connection in order to see the form of the dilaton
couplings induced by the non-Riemannian formulation. Thus, we
interpreted the deviations of the world lines of test particles
from geodesics of the Levi-Civita connection as nonmetricity and
torsion related effects. Finally, we examined the possible spinor
couplings to the model and found that there is no nontrivial
dilaton couplings to Dirac spinors in two dimension. People
interested in nontrivial dilaton couplings to spinors in two
dimensions have to consult \cite{madak04} in which the authors
used the Kaluza-Klein dimensional reduction scheme.

 \section*{Acknowledgement}

The author thanks to D Grumiller and M O Katanaev for their
helpful comments.

\end{document}